\documentclass{aa}

\usepackage{graphicx}
\usepackage{txfonts}
\usepackage{lipsum}
\usepackage{subcaption}
\usepackage{lscape}
\usepackage{placeins}

\usepackage{hyperref}

\begin{document}

   \title{LACHESIS: Robust stellar parameters through Bayesian model averaging of isochrone grids}

   \author{Jos\'{e} I. Vines\inst{1}\corrauth{jose.vines@ucn.cl}
        \and
        Austin T. Ware\inst{2}\thanks{Present address: Department of Physics and Astronomy, University of Notre Dame, Notre Dame, IN 46556, USA}\email{aware2@nd.edu}
    }

   \institute{Instituto de Astronom\'ia, Universidad Cat\'olica del Norte, Angamos 0610, 1270709, Antofagasta, Chile \label{inst1}
        \and
        School of Earth and Space Exploration, Arizona State University, Tempe, AZ 85287, USA \label{inst2}
   }

   \date{Received September 30, 20XX}

  \abstract
   {Accurate stellar parameters underpin much of modern astrophysics, from the characterisation of exoplanetary systems to Galactic archaeology. For isolated field stars, masses and especially ages are most often inferred by comparing observed properties against stellar evolution models, yet these inferences depend on the chosen model grid, and the systematic differences between grids can rival the statistical uncertainties of any single pipeline.}
   {We present \textsc{lachesis}, a \textsc{python} package that determines stellar masses, radii, and ages from photometry and spectroscopy while explicitly accounting for the systematic uncertainties introduced by the choice of stellar evolution model grid.}
   {We interpolate five independent isochrone grids in [Fe/H], log age, and equivalent evolutionary phase conditioned on the data, sample each grid using nested sampling, and combine the resulting per-grid posteriors using Bayesian model averaging, with weights proportional to the Bayesian evidence of each grid.}
   {We validate \textsc{lachesis} against a benchmark of Kepler asteroseismic dwarfs and subgiants. Masses are recovered with a robust scatter of $\sim4\%$ and radii to $\sim2\%$, with the radius negligibly biased by $<1\%$ and the masses carrying a small $\sim-3\%$ systematic. Stellar age is recovered with a robust scatter of $30\%$ that depends strongly on evolutionary state; the grid-to-grid systematic contributes a $\sim9\%$ floor, a subdominant but real term that a single-grid fit omits. Model averaging is better calibrated than selecting a single grid, its credible intervals approach nominal coverage once the reference uncertainty is included, and cover near the nominal rate in injection tests even when the generating model lies outside the ensemble.}
   {By marginalising over an ensemble of stellar model grids, \textsc{lachesis} folds the choice-of-grid systematic directly into the posterior, so the reported uncertainties reflect the choice-of-grid systematic rather than only within-grid data noise, and stay near nominal in injection tests even when the generating model lies outside the ensemble. This delivers homogeneous, well-characterised masses, radii, and ages for exoplanet host characterisation and stellar population studies.}


   \maketitle
   \nolinenumbers

\section{Introduction}

Determining precise and accurate stellar properties remains a central challenge in stellar astrophysics, with wide-ranging implications for fields from Galactic archaeology to exoplanet characterisation. In particular, measuring stellar ages, especially for isolated field stars, is notoriously difficult and no single age-dating technique is universally applicable across the Hertzsprung–Russell diagram \citep[e.g.,][]{soderblom2010}. Among these techniques, isochrone fitting, in which observed stellar properties are compared with predictions from grids of theoretical stellar evolution models along loci of common age or evolutionary phase, remains one of the most widely used and broadly applicable approaches for estimating the masses and ages of field stars. However, degeneracies between stellar parameters, observational uncertainties, the complex morphology of isochrones, and the diversity of physical assumptions in stellar evolution codes make the inference problem highly non-linear and prone to biases.

In an effort to better address these challenges, Bayesian methods have become ubiquitous for stellar parameter inference. Early work introduced a formal Bayesian framework for isochrone fitting, enabling more robust estimates of stellar parameter uncertainties through the computation of posterior probability distributions rather than single-point estimates \citep{joergensen2005}. Subsequent work implemented sampling-based approaches, such as Markov Chain Monte Carlo and nested sampling \citep[e.g.,][]{maxted2015,isochrones}, to improve the flexibility and efficiency of parameter space exploration and incorporated diverse observational constraints into the Bayesian framework \citep[e.g.,][]{lin2018}.

Despite these advances, a key limitation of most current approaches is their reliance on a single set of stellar model grids. Stellar evolution codes differ in their treatment of physical processes such as convection, convective overshooting, microscopic diffusion, rotation, helium enrichment, and chemical composition \citep[e.g.,][]{dotter_dartmouth_2008,bressan_span_2012,dotter_mesa_2016,spada_yalepotsdam_2017,hidalgo_updated_2018}, leading to systematic differences in inferred stellar properties. Systematic uncertainties due to offsets between model grids can be comparable to, or larger than, the statistical uncertainties reported by any given modelling pipeline. Comparisons of inferred stellar properties between widely used model grids suggest systematic uncertainties of $\sim 5\%$ in mass and $\sim 20\%$ in age for main-sequence and subgiant stars \citep{tayar2022}. Although many studies have compared results between stellar model grids or explored the effects of varying model assumptions, the primary purpose is to assess the spread in inferred parameters rather than to assess the relative statistical support the data provide for each model grid \citep[e.g.,][]{southworth2012,maxted2015,silvaaguirre2015,tayar2022}. In other words, uncertainties arising from the choice of model grid are typically treated as external systematic errors rather than accounted for within a unified statistical framework. This holds even for the most recent large-scale efforts. \citet{boin26} train independent neural networks on five evolutionary grids, four of which (BaSTI, MIST, PARSEC, and Dartmouth) we also adopt, to determine ages for 1.3 million stars; but the grids are applied in parallel and their dispersion is reported as systematic rather than combined within a single statistical framework.

Bayesian model averaging (BMA) provides a natural and principled solution to this problem. BMA is a statistical framework used to account for model uncertainty \citep{hoeting1999} and is an extension of Bayesian inference commonly used in the literature. Instead of selecting a single model, or model grid, BMA combines the predictions, or posterior distributions, from multiple models, weighting each model by how well it predicts the data. BMA has been used widely across astrophysics \citep[for a review, see][]{parkinson2013} and recently for the related topic of stellar spectral energy distribution fitting \citep{vines_span_2022}, but it has not yet been used for the purpose of stellar isochrone fitting.

Here we present \textsc{lachesis}\footnote{https://github.com/jvines/LACHESIS} (modeL Averaging for isoChrone cHaractErisation of Stellar propertIeS), a package for applying BMA to stellar isochrone fitting that accounts for model-grid-dependent systematic uncertainties in inferred stellar properties. The methodological difference of the present paper with prior work is that, rather than running a single, or multiple grids separately and quote the spread between them, \textsc{lachesis} weights each grid by its Bayesian evidence and marginalises over the ensemble, so the uncertainty arising from the choice of stellar model grid is folded into a single posterior on stellar mass, radius, and age rather than reported alongside it. By weighting the constraints provided by a suite of stellar model grids, this framework aims to improve the robustness of stellar mass and age estimates and provide uncertainties that more fully reflect model grid choice. This paper is structured as follows: Section~\ref{sec:methods} describes the method. We construct the isochrone grids (Section~\ref{sec:grids}), retrieve and quality-control the photometry (Section~\ref{sec:phot}), select the priors (Section~\ref{sec:priors}), set out the Bayesian Model Averaging framework (Section~\ref{sec:bma}), and describe the nested-sampling implementation (Section~\ref{sec:sampling}); the interpolation coordinate is validated in Appendix~\ref{sec:interp}. Section~\ref{sec:sample} presents the benchmark sample of asteroseismic stars and detached eclipsing binaries used to validate the code. Section~\ref{sec:results} reports the recovery of stellar mass, radius, and age (Section~\ref{sec:recovery}), the model-systematic floor (Section~\ref{sec:floor}), the recovery across evolutionary state (Section~\ref{sec:regime}), the calibration of the reported uncertainties (Section~\ref{sec:errors}), the comparison of model averaging against single-grid selection (Section~\ref{sec:averaging}), and an injection-recovery test of the averaged posterior including a generating model outside the ensemble of grids used in the averaging (Section~\ref{sec:injection}). We discuss the results and their limitations in Section~\ref{sec:disc} and summarise the work and future directions in Section~\ref{sec:conc}.

\section{Isochrone modelling with \textsc{lachesis}}
\label{sec:methods}
\textsc{lachesis} provides an end-to-end interface for automatically fitting up to five different isochrone grids to photometric observations with minimal user input (coordinates and ideally the Gaia DR3 source id, and we allow the user to input their own photometry, and their own stellar parameters encoded as priors) to obtain the Equivalent Evolutionary Phase \citep[EEP;][]{dotter_mesa_2016}, the metallicity, the age of the star, distance, and V-band extinction. In addition to these, \textsc{lachesis} offers the following computed parameters: initial mass, mass, effective temperature, surface gravity, luminosity, and density.

We model each isochrone with 5 + $n$ parameters, as shown in Equation \ref{eq:model}, where $M_{\rm bol}$, $T_{\rm eff}$, and $\log g$ are obtained by trilinear interpolation of the isochrone grid at the sampled EEP ($\mathcal{E}$), log age ($\log\tau$), and [Fe/H] point; ${\rm BC}_i$ is the bolometric correction for band $i$, interpolated from the MIST synthetic photometry tables \citep{choi_mesa_2016}; $\mu$ is the distance modulus $5\log_{10}\!\left(\frac{d}{10\,{\rm pc}}\right)$; and $\sigma_i$ is a photometric excess term added to each photometric measurement, expanded in Equation~\ref{eq:jitter}, where $\sigma_{m, i}$ is the $i$-th measurement uncertainty and $\sigma_{j, i}$ is our added jitter term. The $n$ jitter terms are band-specific, one per photometric measurement, acting as a per-band outlier absorber.

\begin{equation}
\label{eq:model}
    m_i = M_{\rm bol}(\mathcal{E}, \log\tau, {\rm [Fe/H]})
        - {\rm BC}_i(T_{\rm eff}, \log g, {\rm [Fe/H]}, A_V)
        + \mu
\end{equation}

\begin{equation}
\label{eq:jitter}
    \sigma_i^2 = \sigma_{m,i}^2 + \sigma_{j, i}^2
\end{equation}

All five grids deliberately share the same MIST bolometric corrections, both to span the full UV-to-mid-IR band set uniformly and so that the model evidences compare the grids' predicted $(T_{\rm eff}, \log~g, L)$ surfaces on a common photometric scale rather than differences between their native synthetic-photometry libraries. Each grid's effective temperature scale still enters through the corrections, so the comparison isolates interior structure only up to that term. A common numerical metallicity is also not a common physical composition. Each grid adopts its own solar reference and [Fe/H]$\rightarrow Z$ mapping, so the implied initial metal mass fraction at fixed [Fe/H] spans $\sim0.06$~dex across the five grids (a factor $1.15$, from $Z=0.014$ for MIST to $0.016$ for Dartmouth), which we treat as part of the inter-grid systematic that the model average reflects.

\subsection{Model grid construction}
\label{sec:grids}
\textsc{lachesis} fits five isochrone grids simultaneously: MIST
\citep{dotter_mesa_2016}, PARSEC \citep{bressan_span_2012}, Dartmouth
\citep{dotter_dartmouth_2008}, BaSTI \citep{hidalgo_updated_2018}, and YaPSI
\citep{spada_yalepotsdam_2017}. These grids differ in their input physics (opacities,
nuclear reaction rates, convective mixing prescriptions), coverage in mass, age, and metallicity, and in their native parameterisation. MIST provides physics-based EEPs anchored to evolutionary milestones (ZAMS, TAMS, RGB tip, and so on). PARSEC provides evolutionary phase labels that we translate to the MIST EEP scheme via linear interpolation within each phase block. Dartmouth provides its own compact native EEPs. BaSTI and YaPSI ship without EEPs; their isochrones are natively indexed by mass-sorted row number. For YaPSI, we reconstruct physics-based EEPs from the central-hydrogen fraction tabulated in its published tracks, anchoring the primary points to the milestones of \citealt{dotter_mesa_2016} and spacing the rest by the same metric distance. We show the coverage of each isochrone grid in Table \ref{tab:grids}.

To properly fit these different grids coherently, we adopt the EEP \citep{dotter_mesa_2016} as a common evolutionary coordinate. The coordinate plays two distinct roles. As a prior coordinate, the EEP enters the change of variables from the initial-mass function. Consider a density $p(\mathcal{M})$ on the initial mass $\mathcal{M}$, and let $\mathcal{M}(\mathcal{E}, \log\tau, {\rm [Fe/H]})$ be the initial mass returned by the interpolator. At fixed $\log\tau$ and [Fe/H], $\mathcal{M}$ is monotonic and injective in $\mathcal{E}$, so the standard change of variables gives

\begin{equation}
  p(\mathcal{E}) = p\bigl(\mathcal{M}\bigr)
  \;\left|\frac{\partial\mathcal{M}}{\partial\,\mathcal{E}}\right|
  \label{eq:eep_transform}
\end{equation}

With the Jacobian $|\partial\mathcal{M}/\partial\mathcal{E}|$ pre-computed by finite differences along the EEP axis. Equation~\ref{eq:eep_transform} is exact for any monotonic mapping, whether $\mathcal{E}$ is a physics-based EEP or a mass-sorted index, so the prior is correctly normalised in either case. The choice of $p(\mathcal{M})$ is deferred to Section~\ref{sec:priors}.

However, as an interpolation coordinate, the EEP carries a stronger requirement: a fixed $\mathcal{E}$ must correspond to the same fractional evolutionary phase at every age and metallicity, so that interpolating the grid along the $\log\tau$ and [Fe/H] axes blends stars in like evolutionary states. A mass-sorted index does not guarantee this. At fixed mass index, the evolutionary state drifts with age, because the main-sequence turnoff moves to lower mass as the isochrone ages; near the turnoff a single index can label a dwarf on a young isochrone and a subgiant on an old one. Interpolating in age then blends a dwarf with a subgiant and smears the turnoff, precisely where the age is most tightly constrained. Whether this is harmful is set by the native mass sampling: a finely sampled isochrone keeps the phase drift between adjacent indices small and the blend negligible, while a coarse one does not. BaSTI's isochrones are sampled at $\sim2000$ mass points, fine enough that its mass index tracks evolutionary phase smoothly, and its recovered ages agree with the EEP-based grids to within the inter-grid scatter (Section~\ref{sec:floor}); a direct held-out test of the coordinate (Appendix~\ref{sec:interp}) confirms this, and we retain it. YaPSI's native isochrones carry only $\sim70$ mass points, too few to resolve the turnoff. However, the YaPSI tracks tabulate the central hydrogen fraction $X_c$, enabling us to reconstruct EEPs on the \citealt{dotter_mesa_2016} scale.
We anchor four primary points, the zero-age main sequence at $X_c=X_{c,{\rm init}}$, an intermediate point at $X_c=0.3$, the terminal-age main sequence at $X_c\simeq0$, and the base of the red-giant branch at the first $T_{\rm eff}$ minimum past the turnoff, and between consecutive anchors place the integer EEPs by the metric-distance prescription of \citealt{dotter_mesa_2016}, finely enough that adjacent EEPs differ by $\lesssim 0.03$ dex in $\log L$ and $\log T_{\rm eff}$ ($\sim 200$ points per track). At that density the re-indexing reconstructs each track to well below the inter-grid scatter, so the interpolated grid is insensitive to the spacing metric, just as it is to the EEP integers themselves.
Each track is resampled onto the resulting integer EEPs, and the isochrones are rebuilt by interpolating age against initial mass at fixed EEP, so the four-dimensional interpolation runs between homologous evolutionary states.

Each grid is stored as a dense four-dimensional array $\mathcal{G}[i_\text{[Fe/H]},\, i_\text{age},\, i_\mathcal{E},\, k]$ where $k$ indexes the stellar properties. Stellar properties at arbitrary $(\mathcal{E}, \log\tau, \text{[Fe/H]})$ are obtained via trilinear interpolation.

\begin{table*}
\centering
\caption{Isochrone grids used in the Bayesian model averaging with their respective parameter ranges.}
\label{tab:grids}
\begin{tabular}{lcccc}
\hline
Grid & EEPs & [Fe/H] range & $\log\tau$ range & $M_{\rm ini}$ ($M_\odot$) \\
\hline
MIST       & Phase-based   & $-4.0$ to $+0.5$   & $5.0$--$10.3$  & $0.1$--$300$ \\
PARSEC     & Translated    & $-2.0$ to $+0.5$   & $5.0$--$10.25$ & $0.12$--$350$ \\
Dartmouth  & Compact       & $-2.0$ to $+0.5$   & $9.0$--$10.18$ & $0.1$--$2.2$ \\
BaSTI      & Mass-index    & $-2.0$ to $+0.3$   & $9.0$--$10.18$ & $0.1$--$2.4$ \\
YaPSI      & Reconstructed & $-0.75$ to $+0.5$  & $8.7$--$10.18$ & $0.6$--$2.0$ \\
\hline
\end{tabular}
\tablefoot{
The EEP column gives the provenance of each grid's EEPs rather than their values, as the actual numbers are ultimately irrelevant and serve only as the interpolation index.
}
\end{table*}

\subsection{Photometry retrieval and quality control}
\label{sec:phot}
We follow the automatic photometry retrieval procedure of \citet{vines_span_2022}. Briefly, we use \textsc{astroquery} \citep{astroquery} to query the Gaia DR3 archive \citep{gaia_collaboration_gaia_2023} for parallaxes, adopt the geometric distance estimates of \citet{bailer-jones_estimating_2021}, and retrieve archival broad-band photometry from Vizier and MAST. Source identification is anchored to Gaia and propagated through the relevant catalogue crossmatches, allowing us to recover measurements from Gaia DR3, Tycho-2, Johnson UBV, 2MASS, AllWISE, SDSS, Pan-STARRS1, SkyMapper, and Strömgren where available. The benchmark stars carry a median of 15 bands each. We retain only measurements that pass the catalogue quality filters described by \citet{vines_span_2022}, rejecting sources flagged as extended, blended, or otherwise unreliable, and when a band has no reported uncertainty, we assign it a conservative one, $\sigma_{m,\rm missing} = \max_i \sigma_{m,i} + \frac{0.1}{0.921} \approx \max_i \sigma_{m,i} + 0.109\ \text{mag}$, i.e.\ the largest reported uncertainty among the star's other bands plus the magnitude equivalent of a $10\%$ flux error.

We extend the photometric quality control of \citet{vines_span_2022} with an iterative outlier-screening step. We evaluate the MIST bolometric correction tables \citep{choi_mesa_2016} at each trial effective temperature, which accounts for the filter-specific zero-point differences between photometric systems. We first fit this model to the Gaia photometry alone, then iteratively promote the next-best-fitting band into the trusted set, refitting at each step, until no remaining band achieves residuals below $5\sigma$. Bands that are never promoted are flagged as potentially problematic, most often because of saturation, blending, or an erroneous crossmatch. These flags are reported to the user but do not automatically remove the affected bands from the fit. We impose an error floor of 0.05~mag to prevent unrealistically small catalogue uncertainties from producing spurious outlier detections. When no Gaia photometry is available we fall back to iterative worst-outlier removal instead. Figure \ref{fig:bbody} illustrates the procedure.

Finally, \textsc{lachesis} offers an SED plot with a blackbody fit to the photometry so the user can eyeball the retrieved magnitudes and manually remove them before continuing with the fit. We show an example in Figure~\ref{fig:bbody2}.

\subsection{Prior selection}
\label{sec:priors}
The five main parameters sampled by \textsc{lachesis} are the EEP, $\log\tau$, [Fe/H], distance, and V-band extinction A$_{\rm V}$. Here we describe the priors used for these parameters. We write $\mathcal{U}(a, b)$ and $\mathcal{N}(\mu, \sigma)$ for uniform and normal distributions.

\subsubsection{EEP prior}
\label{sec:eep_prior}
Following \citet{isochrones}, we use the \citet{chabrier_galactic_2003} initial mass function (IMF), denoted by $\xi$ in Equation \ref{eq:eep-prior}, weighted by the Jacobian of the mass-EEP transformation; thus the prior distribution for the EEP becomes

\begin{equation}
\label{eq:eep-prior}
    P(\mathcal{E})\propto\xi(M_{\rm ini})\left|\frac{d M_{\rm ini}}{d \mathcal{E}}\right|
\end{equation}

We also offer the IMFs by \citet{salpeter_luminosity_1955} and \citet{kroupa_variation_2001} configurable by the user.

\subsubsection{Age}
\label{sec:age_prior}
The age has a log-uniform prior between 100 Myr and 20 Gyr, resulting in a $\mathcal{U}(8, 10.3)$ prior for $\log\tau$. The upper bound is deliberately looser than the age of the Universe so as to not truncate old-star posteriors; in practice the age is bounded by each grid's coverage (Table~\ref{tab:grids}). We also offer the prior used by \citet{isochrones}, uniform in linear age, $P(\log\tau) \propto 10^{\log\tau}$.

\subsubsection{Metallicity}
\label{sec:met_prior}
For metallicity, we offer different types of priors: population priors, spectroscopic priors, and a $\mathcal{U}(-2, 0.5)$ prior, which covers the majority of the range from Table \ref{tab:grids}. The two population priors are a $\mathcal{N}(-0.125, 0.234)$, which is an empirical prior drawn from RAVE DR5 \citep{RAVE}, and a two Gaussian mixture disc plus halo model from \citet{casagrande_new_2011}. Finally, the spectroscopic priors depend on the data availability in different catalogues: Hypatia \citep{hypatia}, PASTEL \citep{Soubiran2016}, RAVE \citep{steinmetz_sixth_2020}, GALAH \citep{buder_galah_2021}, APOGEE \citep{abdurrouf_seventeenth_2022}, and LAMOST \citep{2026yCat.5162....0L}. These catalogues are queried sequentially in the following order: Hypatia > PASTEL > APOGEE > GALAH > RAVE > LAMOST. We query Hypatia first because it is a multi-study compilation placed on a common solar scale (here the \cite{asplund09} composition). Its [Fe/H] aggregates many independent high-resolution determinations and its uncertainty reflects the spread between those studies, a more realistic error than the per-catalogue internal estimates, which underrepresent the method-to-method systematics. The remaining catalogues act as a fallback when a star is absent from Hypatia. For all the catalogues discussed here, the prior is always $\mathcal{N}(\mu_{[Fe/H]}, \sigma_{[Fe/H]})$ truncated to the grid coverage; for Hypatia specifically, $\mu_{[Fe/H]}$ is the median of the measurements, and $\sigma_{[Fe/H]}$ is the inter-study dispersion with a floor of 0.1~dex.

\subsubsection{Distance and Extinction}
\label{sec:dist_prior}
The distance prior is queried from the \citet{bailer-jones_estimating_2021} eDR3 distance catalogue, and the extinction prior is first queried through the Bayestar 3D dustmap \citep{2019ApJ...887...93G}, if nothing is found, then we use a $\mathcal{U}(0, a_v)$ prior, where $a_v$ corresponds to the maximum in line-of-sight from the SFD dustmaps \citep{2011ApJ...737..103S}, both through the \textsc{dustmaps python} package \citep{2018JOSS....3..695M}. For stars with distances $\leq70$ pc we fix the extinction to zero, though this can be overridden by the user. Extinction is applied through the MIST bolometric corrections assuming a \citet{fitzpatrick99} law with $R_V = 3.1$.

\subsubsection{Photometric jitter}
The excess-noise term has a log-uniform prior, $\log_{10}\sigma_j \sim \mathcal{U}(-3,-0.3)$, corresponding to $\sigma_j \in [10^{-3},0.5]~\mathrm{mag}$. This broad prior allows the noise floor to remain effectively off unless the data require additional inflation.

\subsection{Bayesian model averaging}
\label{sec:bma}
Recall Bayes' theorem which relates a model $\mathcal{M}$, with parameters $\bar{\theta}$, and data $D$

\begin{equation}
    P(\bar\theta|D,\mathcal{M}) = \frac{P(D|\bar\theta,\mathcal{M})P(\bar\theta|\mathcal{M})}{P(D|\mathcal{M})}
\end{equation}

Where $P(\bar\theta|\mathcal{M})$ is the prior probability distribution for model $\mathcal{M}$, $P(D|\bar\theta,\mathcal{M})$ is the likelihood of the data given the model and some parameter vector $\bar\theta$, $P(\bar\theta|D,\mathcal{M})$ is the posterior probability distribution of the parameters, and $P(D|\mathcal{M})$ the Bayesian evidence.

In the single-model inference case, the evidence serves as nothing more than a normalisation factor, ensuring the posterior probability distribution normalises to unity. However, when considering multiple models, the evidence becomes the cornerstone of the BMA method.

Assuming there is a prior distribution over the set of all considered models, if there is a probability mass function over all models with values $P(\mathcal{M}_l)$ for $l=1,...,K$, then we can use Bayes' theorem to derive the posterior model probabilities given by the data

\begin{equation}
\label{eq:model-bayes}
    P(\mathcal{M}_l|D)=\frac{P(D|\mathcal{M}_l)P(\mathcal{M}_l)}{\sum_{m=1}^KP(D|\mathcal{M}_m)P(\mathcal{M}_m)}
\end{equation}

Naturally, the probabilities given by Equation \ref{eq:model-bayes} can be used for model selection, but a more interesting extension of this methodology appears when a parameter $\Theta$ is present in all trial models. Then, its marginal posterior distribution across all models becomes

\begin{equation}
\label{eq:bma-parameter}
    P(\Theta|D)=\sum_{l=1}^KP(\Theta|D,\mathcal{M}_l)P(\mathcal{M}_l|D)
\end{equation}

which is effectively an average of all posterior distributions weighted by each posterior model probability. Herein lies the advantage of BMA, a direct combination of models to obtain parameter estimates.

For this work we adopt a uniform prior over the model set, simplifying Equation~\ref{eq:model-bayes} to normalising the evidence of each model by the sum of all evidences. This is a convention rather than an uninformative prior: it treats near-duplicate grids as independent votes, so an ensemble weighted towards one family of codes would shift the average. Our five grids come from independent evolution codes, which mitigates but does not eliminate this dependence.

Because each grid is interpolated over its own coverage in mass, age, and metallicity, the raw Bayesian evidence returned by nested sampling depends not only on how well a grid fits the data but also on the extent of its coverage. A grid whose tracks span a narrower mass range or whose evolutionary index is more finely sampled would otherwise receive an arbitrary boost or penalty in the model weights. To place the grids on equal footing we normalise every grid's prior to a common scale before forming the weights. The IMF prior is normalised over a single reference mass range shared by all grids, and the log-age and metallicity priors are taken flat on common intervals (i.e. the union of grid coverages). The evidence then measures the marginal likelihood of the data under a prior that is identical across grids, so the weights reflect fit quality rather than the extent of a grid's tracks. This normalisation shifts each grid's log-evidence by a constant and leaves the per-grid posteriors unchanged.

In \textsc{lachesis} we adopt a Gaussian likelihood (Equation \ref{eq:likelihood}) with Equation \ref{eq:model} as the mean, where $\bar{m_i}$ is the observed magnitude for filter $i$, $m_{i,\mathcal{M}}$ is the predicted apparent magnitude from isochrone grid $\mathcal{M}$, and $\sigma_i$ is the total per-band uncertainty of Equation~\ref{eq:jitter}. The same likelihood accepts non-photometric observables. Any quantity the grids predict, such as a spectroscopic effective temperature, surface gravity, or metallicity, or a measured luminosity or radius, enters Equation~\ref{eq:likelihood} as a further Gaussian term and can be fit together with the broad-band magnitudes or in place of them; the spectroscopy-only DEBCat fits of Section~\ref{sec:results} ($T_{\rm eff}$ and [Fe/H], no photometry) are an example of the latter.

\begin{equation}
\label{eq:likelihood}
    \mathcal{L} = \prod_{i=1}^N\frac{1}{\sqrt{2\pi\sigma_i^2}}\exp\left(-\frac{1}{2}\left(\frac{\bar{m_i}-m_{i,\mathcal{M}}}{\sigma_i}\right)^2\right)
\end{equation}

\subsection{Posterior sampling}
\label{sec:sampling}
The Bayesian evidence is the cornerstone of BMA, but it has the disadvantage that it is usually hard to calculate numerically or analytically, the latter being impossible in most practical applications. Hence why we use the nested sampling algorithm \citep{Skilling2004, skilling_nested_2006} through \textsc{dynesty} \citep{Speagle2019}. Nested sampling is designed to compute the Bayesian evidence while still producing posterior distributions.

Two notes on implementation: first, because the EEP prior (Equation \ref{eq:eep-prior}) depends on the grid interpolation at each proposed point, it cannot be encoded in the nested sampling prior transform and is instead evaluated as an additive term in the log-likelihood. Second, a grid does not necessarily contribute to every star. It is excluded a priori when an informative [Fe/H] prior falls outside its metallicity coverage, or when the star is evolved beyond the grid's reach, and a posteriori when its [Fe/H] posterior rails against the grid's axis boundary (more than half the samples within a tolerance of the edge), which signals a star outside the grid's coverage.

We sample each grid with \textsc{dynesty}'s static nested sampler using 500 live points, multi-ellipsoidal bounds, and random-walk proposals, stopping when the estimated remaining evidence satisfies $\Delta\ln Z<0.01$.

\section{Benchmark sample}
\label{sec:sample}
\begin{figure}
    \centering
    \includegraphics[width=0.9\linewidth]{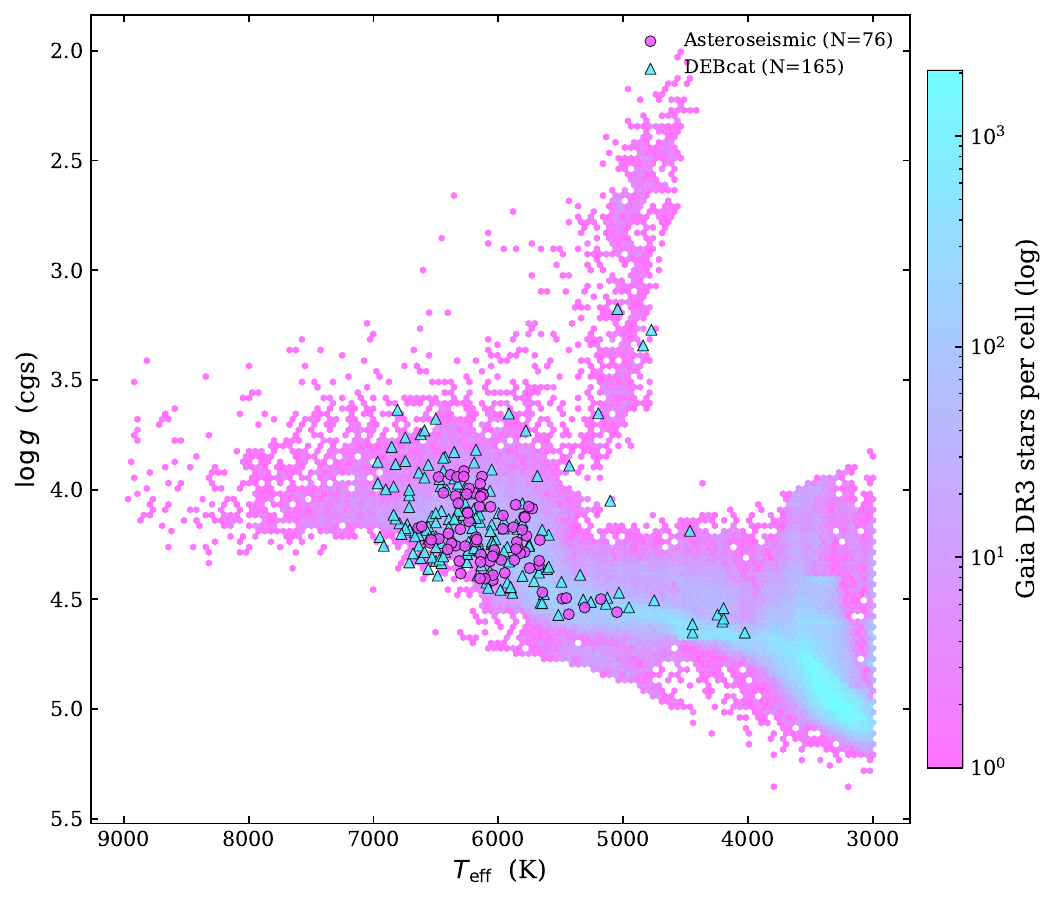}
    \caption{Kiel diagram of the 241 benchmark stars, divided by provenance (asteroseismic shown with circles and detached eclipsing binaries shown with triangles), over the number density of nearby Gaia DR3 field stars (logarithmic colour scale, GSP-Phot $T_{\rm eff}$ and $\log g$). The plotted parameters are literature values: $\log g$ follows from the benchmark mass and radius, $\log g = \log_{10}(M/R^2) + 4.437$, and $T_{\rm eff}$ is taken from the source catalogues for the eclipsing binaries and curated stars, and from \citet{silvaaguirre17} and \citet{mathur17} for the asteroseismic targets. The sample covers the FGK main-sequence and subgiant branch.}
    \label{fig:hrd}
\end{figure}

The validation sample we used is composed of 241 stars (155 F-type, 65 G-type, 21 K-type) main-sequence and subgiant branch stars. We assembled the sample from two independent sub-samples. A sample of stars with asteroseismic observations (N = 76) from the Kepler LEGACY sample \citep{lund17, silvaaguirre17} and the \citet{davies16} planet hosts, compiled by \citet{betrisey26}, which serves to anchor the mass and age of the stars and are dominated by F and G dwarfs (47 F, 27 G, 2 K). The detached eclipsing binaries catalogue \citep[$N=165$;][]{southworth15} spans the full FGK range (108 F, 38 G, 19 K) with dynamical masses; giant and extragalactic systems with no usable Gaia parallax, were excluded. We summarise the benchmark provenance in Table~\ref{tab:sample}, show the sample in the $T_{\rm eff}$-$\log g$ plane in Figure~\ref{fig:hrd}, and give the complete table of stars and their published parameters in Table~\ref{tab:stars}. The combined-prior and metallicity recovery results use the 192 stars with a spectroscopic [Fe/H]; the no-prior photometric fits are run on all 241, with the DEBCat systems included to characterise the single-star blending bias.

Because \textsc{lachesis} fits each star as a single, we remove unresolved astrometric binaries from the asteroseismic sample: 14 stars with Gaia DR3 ${\rm RUWE} > 1.4$, whose unseen companion inflates the photometric luminosity and the radius and age derived from it, leaving 76 out of the original 90. The eclipsing binaries are kept deliberately, both as the dynamical-mass anchor and as a binary-contamination stress test for the single-star photometric radius.

\begin{table}
\centering
\caption{\textsc{lachesis} benchmark sample.}
\label{tab:sample}
\begin{tabular}{lrcc}
\hline\hline
Sub-sample & $N$ & Reference & Anchor \\
\hline
Asteroseismic  & 76  & \citet{betrisey26}
             & Mass and age \\
Detached EB    & 165 & \citet{southworth15}
             & Mass \\
\hline
\end{tabular}
\end{table}

\begin{table*}
  \centering
  \caption{Benchmark stars and their published parameters.}
  \label{tab:stars}
  \begin{tabular}{llcccccc}
  \hline\hline
  Star & Sample & $T_{\rm eff}$ (K) & [Fe/H] & $M$ ($M_\odot$) & $R$ ($R_\odot$) & Age (Gyr) & SpT \\
  \hline
  KIC 2837475 & asteroseismic & $6635$\tablefootmark{a} & $-0.05\pm0.01$ & $1.421\pm0.039$ & $1.625\pm0.016$ & $1.741\pm0.372$ & F \\
  KIC 3427720 & asteroseismic & $6080$\tablefootmark{a} & $-0.02\pm0.01$ & $1.151\pm0.026$ & $1.133\pm0.009$ & $2.244\pm0.096$ & F \\
  KIC 3456181 & asteroseismic & $6331$\tablefootmark{a} & $-0.22\pm0.01$ & $1.342\pm0.049$ & $2.071\pm0.025$ & $2.815\pm0.191$ & F \\
  \multicolumn{8}{c}{$\cdots$} \\
  AD Boo & DEBCat & $6577\pm121$ & $+0.10\pm0.15$ & $1.414\pm0.009$ & $1.612\pm0.014$ & -- & F \\
  AI Phe & DEBCat & $6259\pm42$ & $-0.14\pm0.10$ & $1.194\pm0.001$ & $1.805\pm0.002$ & -- & F \\
  AL Ari & DEBCat & $6368\pm59$ & $-0.42\pm0.08$ & $1.164\pm0.001$ & $1.372\pm0.009$ & -- & F \\
  \hline
  \end{tabular}
  \tablefoot{
  First three stars per sub-sample shown; the full catalogue of 241 stars is available as supplementary material.\\
  \tablefoottext{a}{Asteroseismic-sample $T_{\rm eff}$ values were derived from the published luminosity and radius using $T_{\rm eff}=T_\odot\,(L/L_\odot)^{1/4}\,(R/R_\odot)^{-1/2}$; no $T_{\rm eff}$ uncertainty is propagated because the parent compilation \citep{betrisey26} does not quote one independent of the $L$ and $R$ posteriors.}
  }
\end{table*}

\section{Results}
\label{sec:results}

We tested \textsc{lachesis} in three ways. First, we performed a photometric fit without priors (hereafter No-prior) of the 76 asteroseismic Kepler LEGACY single stars, to measure the intrinsic recovery of the stellar parameters from photometry alone. The second experiment (Combined-prior) adds a spectroscopic [Fe/H] prior, sourced through the catalogue chain of Section~\ref{sec:priors}. For each star, we adopt the Hypatia Catalog \citep{hypatia} value when it is available, taking the median over Hypatia's compiled high-resolution determinations with their inter-study dispersion as the uncertainty, on the \citet{asplund09} solar scale. For stars absent from Hypatia we fall back to the individual surveys, using a single value as published and, where two or more surveys overlap, combining them by inverse-variance weighting with a Birge-ratio inflation of the uncertainty when they disagree by more than their quoted errors \citep[$\chi^2/\mathrm{dof}>1$;][]{birge1932, jofre14}. All adopted uncertainties are floored at 0.1 dex to absorb residual survey-to-survey systematics. The third experiment fits, using no photometry, the 102 DEBCat primaries that carry a published spectroscopic [Fe/H] and $T_{\rm eff}$, because for the vast majority the broad-band SED is the blended light of both components and is unsuitable for isochrone interpolation. This third spectroscopy-only case serves only as a cross-check of the mass scale, and unless stated otherwise, all results below refer to the Combined-prior and No-prior cases. Figure~\ref{fig:blending} shows the \textsc{lachesis} results for the DEBCat stars for the No-prior case. The single-star photometric fit overestimates the radii of these systems by $\sim$30\%, the signature of the unresolved companion's flux.

Throughout, we quantify recovery by the fractional residual $\Delta x / x = (x_{\rm L} - x_{\rm lit}) / x_{\rm lit}$ and separate accuracy from precision. Accuracy is the signed median residual (the bias) with a bootstrap $95\%$ confidence interval. Precision is reported as two scales of scatter, the robust normalised median absolute deviation NMAD $=1.4826\times{\rm median}(|\Delta x / x - {\rm median}(\Delta x / x)|)$, which recovers the standard deviation for a Gaussian but is insensitive to outliers, and the root-mean-square error ${\rm RMSE}=[N^{-1}\sum_i(\Delta x_i / x_i)^2]^{1/2}$, which captures the full distribution including its tails. To test whether the reported uncertainties are honest we use the pull, $(x_{\rm L} - x_{\rm lit}) / \sqrt{\sigma_{\rm post}^2 + \sigma_{\rm lit}^2}$, with $\sigma_{\rm post}$ the half-width of the $68\%$ credible interval and $\sigma_{\rm lit}$ the reference uncertainty. Its robust width (NMAD) measures whether the residuals match the stated uncertainties. A complete error budget gives unit width, larger values mean overconfidence and smaller values conservative errors. The coverage $C_{68}$ and $C_{95}$ is the fraction of stars whose reference value falls inside the equal-tailed $68\%$ and $95\%$ credible intervals, and should equal $0.68$ and $0.95$ for calibrated posteriors; we also report the probability integral transform (PIT).

The benchmark methods carry their own intrinsic-scatter floors. Grid modelling of individual Kepler oscillation frequencies recovers mass, radius, and age to roughly $5\%$, $2\%$, and $15\%$, respectively \citep{silvaaguirre17, tayar2022}, while detached eclipsing binaries deliver dynamical masses and radii to $1\%$ and $2\%$ \citep{torres10, southworth15} but no independent age. One asymmetry follows. The asteroseismic parameters are model-based, derived from the individual oscillation frequencies that constrain the stellar interior, and are the most reliable reference available; the comparison asks whether \textsc{lachesis} can reproduce them from photometry alone. The DEBCat masses are dynamical and model-independent, but the blended binary photometry restricts us to a spectroscopy-only fit from $T_{\rm eff}$ and [Fe/H], which constrain the mass only weakly, so they serve as a limited cross-check rather than an independent calibration of the mass scale.

\subsection{Parameter recovery}
\label{sec:recovery}

\begin{figure*}
    \centering
    \includegraphics[width=0.9\linewidth]{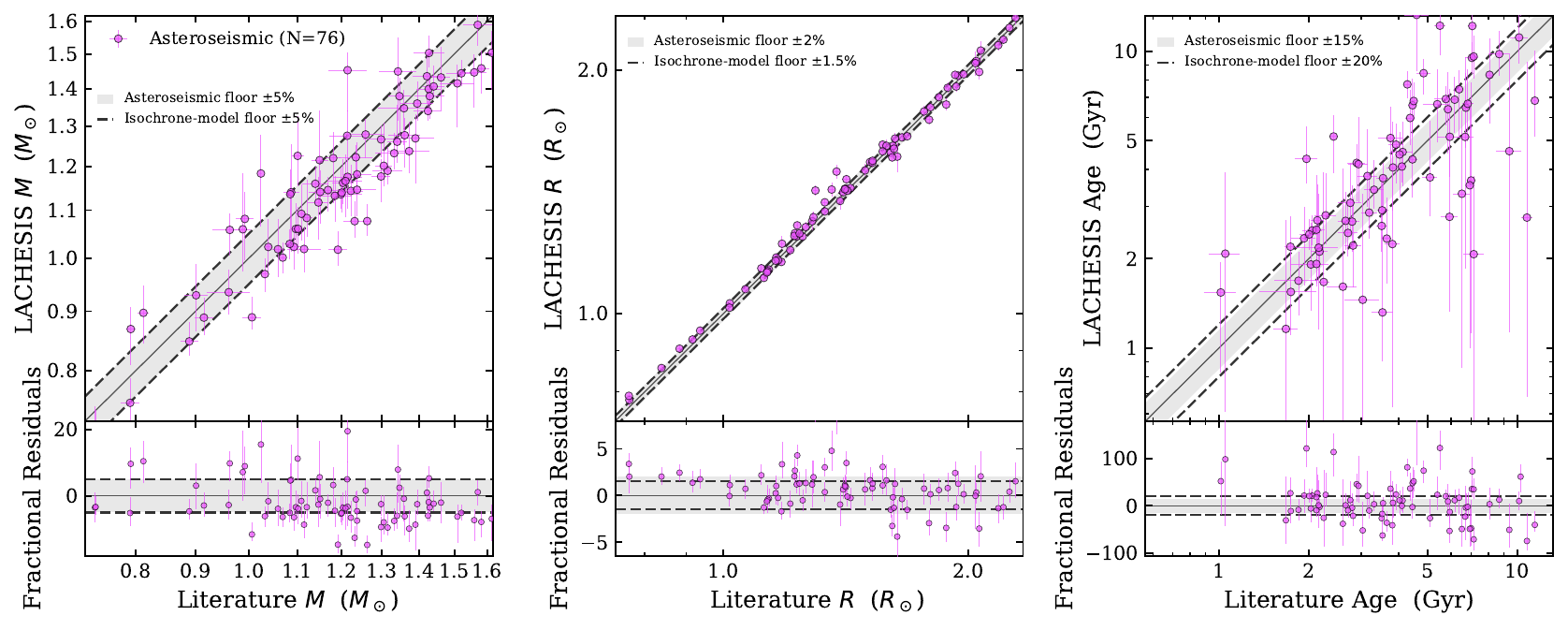}
    \caption{Combined-prior recovery of mass, radius, and age against the asteroseismic benchmarks $N=76$; shaded and dashed bands mark the reference and isochrone-model floors. Mass and radius recover within the floor, age is limited by the photometry with a subdominant model floor.}
    \label{fig:recovery}
\end{figure*}

In the Combined-prior case, \textsc{lachesis} recovers stellar mass and radius with a robust (NMAD) fractional scatter of $4.3\%$ and $1.6\%$ and a full RMSE of $6.9\%$ and $2.7\%$ against the asteroseismic benchmarks (Figure~\ref{fig:recovery}, Table~\ref{tab:calib}). The robust scatter is comparable to the literature grid-modelling floors (Section~\ref{sec:floor}), while the larger RMSE reflects a heavy tail of outlier stars rather than the typical star. The model-systematic component of this scatter, isolated as the inter-grid dispersion in Section~\ref{sec:floor}, is $\sim1.7\%$ in mass and $\sim0.3\%$ in radius and sits well below the floors; the remainder is the data-driven photometric error. Age recovery is limited mainly by the weak photometric leverage on the interior clock, with the inter-grid model floor a subdominant $\sim9\%$ term (Section~\ref{sec:floor}).

\begin{table*}
\centering
\caption{Recovery accuracy, scatter, and calibration against the $N=76$ asteroseismic benchmarks.}
\label{tab:calib}
\begin{tabular}{llccccc}
\hline\hline
Run & Param. & Bias [95\% CI] (\%) & NMAD (\%) & RMSE (\%) & $C_{68}$ / $C_{95}$ & Pull width [95\% CI] \\
\hline
Combined-prior & Mass   & $-3.4\,[-4.4,-2.3]$ & $4.3$ & $6.9$ & $0.51 / 0.76$ & $0.89\,[0.66,1.40]$ \\
         & Radius & $+0.7\,[+0.2,+1.2]$ & $1.6$ & $2.7$ & $0.51 / 0.75$ & $1.08\,[0.78,1.54]$ \\
         & Age    & $+5.7\,[-3.5,+15.0]$ & $32$ & $48$ & $0.65 / 0.90$ & $1.00\,[0.79,1.44]$ \\
\hline
No-prior & Mass   & $+1.5\,[-0.4,+2.7]$ & $6.2$ & $8.8$ & $0.72 / 0.92$ & $0.76\,[0.53,0.96]$ \\
         & Radius & $+0.4\,[-0.0,+0.9]$ & $1.3$ & $2.7$ & $0.61 / 0.83$ & $0.78\,[0.62,1.40]$ \\
         & Age    & $-16.4\,[-22.0,-9.9]$ & $38$ & $61$ & $0.76 / 0.97$ & $0.89\,[0.56,1.06]$ \\
\hline\hline
\end{tabular}
\tablefoot{
Bias is the signed median fractional residual with a bootstrap $95\%$ confidence interval; NMAD is the robust scatter; RMSE is the full scatter; $C_{68}$ and $C_{95}$ are the empirical coverage of the $68\%$ and $95\%$ credible intervals (seismic value treated as exact); the pull width is the robust width of the error-aware pull distribution with its bootstrap $95\%$ interval.
}
\end{table*}

\begin{figure}
    \centering
    \includegraphics[width=1\linewidth]{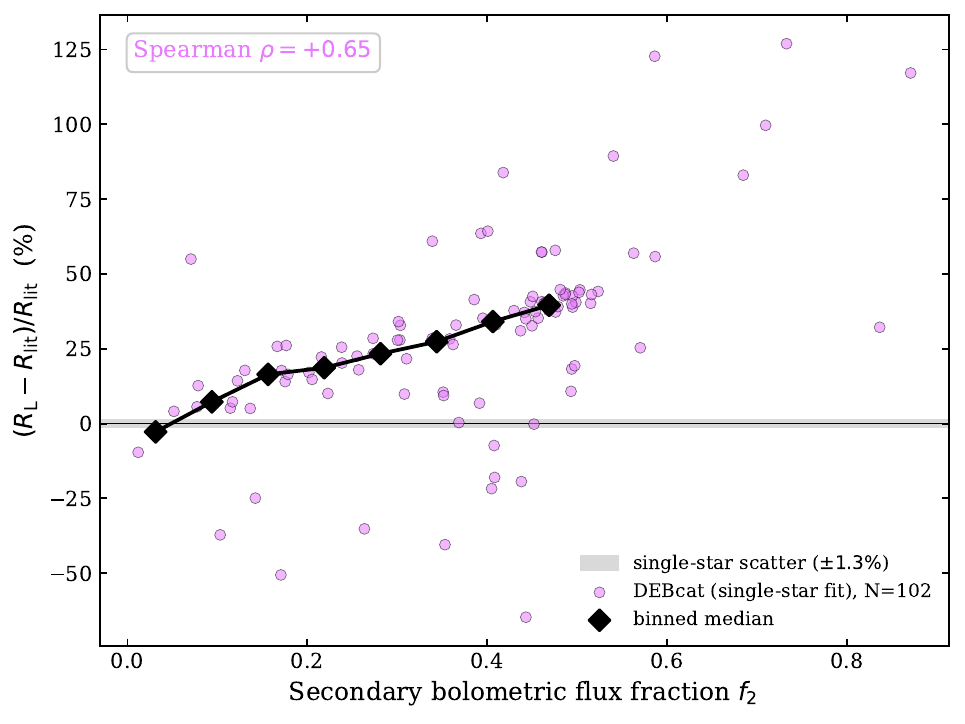}
    \caption{DEBCat radius bias versus the secondary's bolometric flux fraction
  $f_2 \equiv R_2^2 T_2^4/(R_1^2 T_1^4 + R_2^2 T_2^4)$. The single-star photometric
  radius residual (magenta, $N=102$; binned median in black diamonds) is consistent with zero for faint companions and climbs to $\sim\!+30\%$ as the system approaches equal luminosity (Spearman $\rho=+0.65$). The radius outliers are therefore a binarity artefact.}
    \label{fig:blending}
\end{figure}

Against the asteroseismic standard, the spectroscopic [Fe/H] prior shifts the recovery and improves the precision, but without affecting the accuracy. The median age is shifted from $-16.4\%$ (No-prior) to $5.7\%$ (Combined-prior), and the mass bias moves from $1.5\%$ to $-3.4\%$. However, the per-star absolute error is statistically unchanged in all three parameters (Table~\ref{tab:prior}), according to a Wilcoxon signed-rank test on the paired per-star differences between the two runs, which tests whether their median is zero; $p<0.05$ would indicate a significant change. The real gain from the spectroscopic prior is precision. Pinning the [Fe/H] tightens the per-star posteriors by $\sim40\%$ in mass and age, and $\sim10\%$ in radius. The spectroscopy-only fit of the DEBCat primaries marks the limit of the method without photometry: from $T_{\rm eff}$ and [Fe/H] alone the mass is only weakly constrained, and the recovery is biased lower by $6.7\%$, increasingly so above $1.2M_\odot$, reflecting the IMF prior dominating an under-constrained posterior rather than a clean measure of the model mass systematic.

\begin{table}
\centering
\caption{Effect of the [Fe/H] prior on $N=76$ asteroseismic stars.}
\label{tab:prior}
\begin{tabular}{lcccc}
\hline\hline
Parameter & Accuracy & $p$ & Precision & $p$ \\
\hline
Mass (\%)            & $5.0/4.5$   & $0.66$     & $4.8/8.2$   & $<10^{-6}$ \\
Radius (\%)          & $1.3/1.2$   & $0.90$     & $1.2/1.3$   & $<10^{-6}$ \\
Age (\%)             & $22.2/24.7$ & $0.16$     & $28.0/45.6$ & $<10^{-6}$ \\
${\rm [Fe/H]}$ (dex) & $0.02/0.17$ & $<10^{-6}$ & $0.07/0.19$ & $<10^{-6}$ \\
\hline\hline
\end{tabular}
\tablefoot{
Each cell gives Combined-prior\,/\,No-prior. Accuracy is the median absolute residual against the reference, precision the median posterior width, and $p$ a Wilcoxon test on the per-star change.
}
\end{table}

Radius is the best-recovered parameter, with a Combined-prior $1.3\%$ median error (Table~\ref{tab:prior}) and a $0.7\%$ bias (Table~\ref{tab:calib}), below the isochrone model floor. The single-star fit of the blended DEBCat photometry instead overestimates the radius by up to $\sim30\%$ in proportion to the companion's light (Figure~\ref{fig:blending}), so the DEBCat sample enter only as the stress test. The Combined-prior age bias is $5.7\%$, with a per-star median absolute error of $\sim22\%$; the reported posterior is wider, $\sim28\%$ (Table~\ref{tab:prior}).

Metallicity is the one sampled parameter that photometry barely constrains. In the No-prior case, \textsc{lachesis} recovers it to only 0.2 dex (median absolute residual against the spectroscopic value over the $N=192$ photometric fits with a reference), biased $0.13$ dex metal-rich, the broad-band colours preferring a higher metallicity than the spectroscopy. The spectroscopic prior pins it to $0.02$ dex, the Combined-prior fit then propagates into the recovered age and mass.

\subsection{The model-systematic floor}
\label{sec:floor}
\begin{figure}
    \centering
    \includegraphics[width=0.9\linewidth]{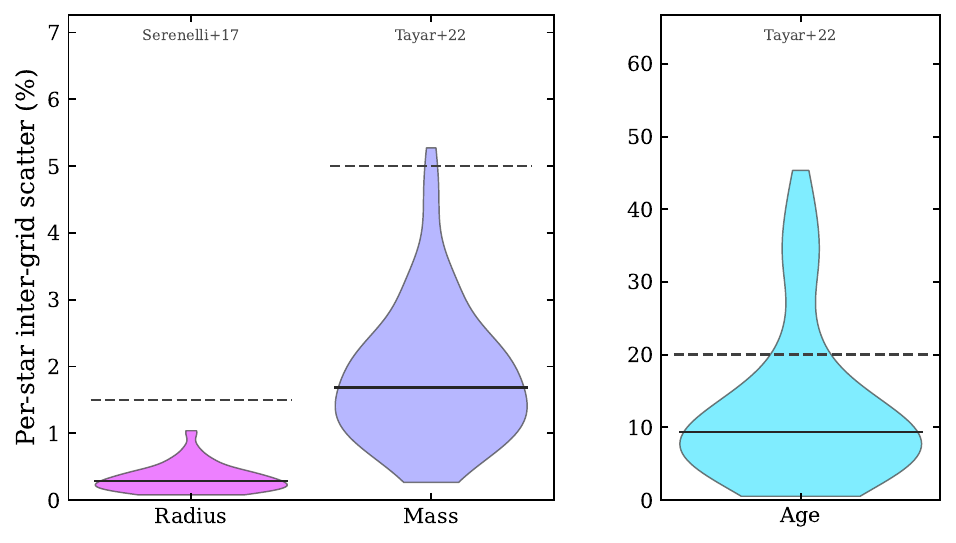}
    \caption{Violin plots with the per-star inter-grid scatter, the dispersion of the five grids at fixed input, for radius, mass, and age (asteroseismic sample). Solid lines are medians, dashed lines the literature systematic floors (radius, \citealt{serenelli17}; mass and age, \citealt{tayar2022}); the scatter runs $\sim0.3\%$ (radius) to $\sim9\%$ (age), below every floor.}
    \label{fig:gridscatter}
\end{figure}

The median per-star inter-grid scatter is $\sim0.3\%$ in radius, $\sim1.7\%$ in mass and $\sim9\%$ in age (Figure~\ref{fig:gridscatter}). This inter-grid scatter is the standard deviation of the five grid medians at fixed input, a model-only quantity that is immune to the truth-scale circularity of comparing against the asteroseismic ages. Because it swaps the grid at fixed input and is blind to any systematic the five grids share, it is a lower bound on the model-dependent floor rather than an absolute lower bound on stellar-model uncertainty. It accordingly sits below the published floors of $1.2-1.6\%$ (radius), $3-4.4\%$ (mass), and $12-28\%$ (age) from \citet{serenelli17} and $5\%$ (mass) and $20\%$ (age) from \citet{tayar2022}, which additionally fold in the input-data, stellar-code, and pipeline terms our comparison holds fixed. The five grids agree on age to better than $10\%$ with no grid dissenting; dropping any one leaves the scatter essentially unchanged, so the $\sim9\%$ is a genuine ensemble property rather than an artefact of a single discrepant grid.

\begin{figure}
    \centering
    \includegraphics[width=0.9\linewidth]{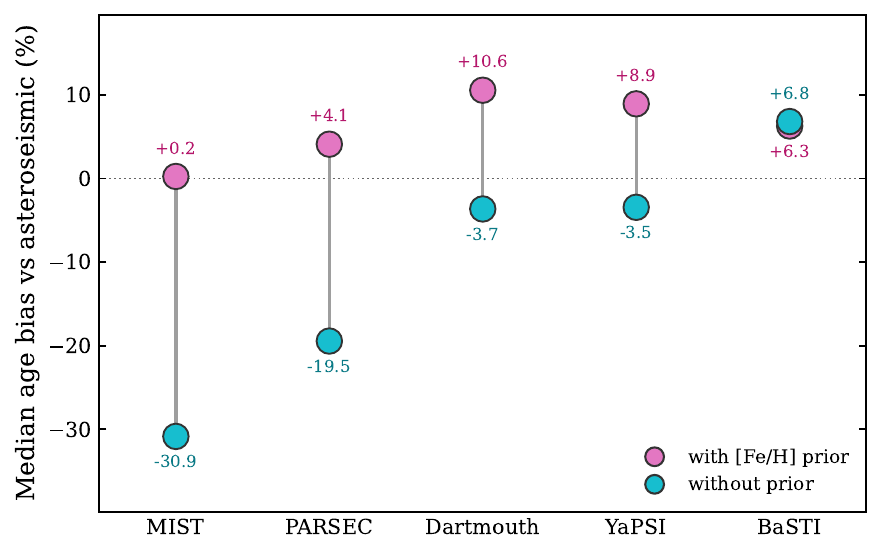}
    \caption{Per-grid median age bias against the asteroseismic reference (76 stars), Combined-prior (pink) and No-prior (cyan), joined per grid and ordered by swing. The prior clusters the grids in a tight 0.2 to 10.6\% band; without it they fan out, MIST reverting to $-31\%$ while BaSTI is unchanged, so no grid's near-zero bias is intrinsic.}
    \label{fig:age-bias}
\end{figure}

The model floor is real but subdominant. A mixture-variance decomposition of the averaged age posterior, ${\rm Var}(\theta|D)=\sum_k w_k {\rm Var}(\theta|D,\mathcal{M}_k)+\sum_k w_k(\mu_k - \mu)^2$, puts the between-grid term at only $\sim11\%$ of the total age variance (median $0.09$); the within-grid data term dominates. The $\sim9\%$ inter-grid dispersion therefore sets a floor beneath the $\sim22-28\%$ per-star age uncertainty rather than dominating it. The metallicity prior tightens the age posterior (median width $45.6\%\rightarrow28.0\%$, Table~\ref{tab:prior}) while leaving the accuracy against the reference statistically unchanged ($24.7\rightarrow22.2\%$, $p=0.16$), because the residual error is set by the model floor and the main-sequence age degeneracy, not by [Fe/H] knowledge. The accompanying $5.7\%$ median age bias is a property of the model set rather than an inference artefact; it is worst for the tightest posteriors (Spearman $\rho=-0.63$ between posterior width and signed bias, the opposite of a posterior reverting to the prior). Since the asteroseismic reference is itself grid-model-derived, it is the inter-grid scatter, not the bias zero-point, that is the absolute statement about isochrone ages. We caution against reading any single grid as the accurate one. The near-zero Combined-prior MIST bias (Figure~\ref{fig:age-bias}) is incidental and changes to $-31\%$ in the No-prior case, while BaSTI barely moves.

No single grid dominates the average. Under the common-scale normalisation (Section~\ref{sec:bma}) the mean model weights are spread evenly across the ensemble, from $\sim0.18$ (PARSEC, Dartmouth, YaPSI) to $\sim0.27$ (MIST), with BaSTI at $0.20$. The evidence measures how well each grid reproduces the photometric surface on the shared bolometric-correction scale, which need not track the accuracy of the derived parameters. BaSTI returns the most accurate masses ($4.8\%$) and radii ($1.1\%$) of the ensemble while carrying only a middling weight. Because all five grids are evaluated on the same MIST bolometric corrections (Section~\ref{sec:methods}), the weights reflect a consistent comparison on a common photometric scale and do not privilege any one grid.

\subsection{Recovery across the evolutionary state}
\label{sec:regime}

\begin{figure}
    \centering
    \includegraphics[width=0.9\linewidth]{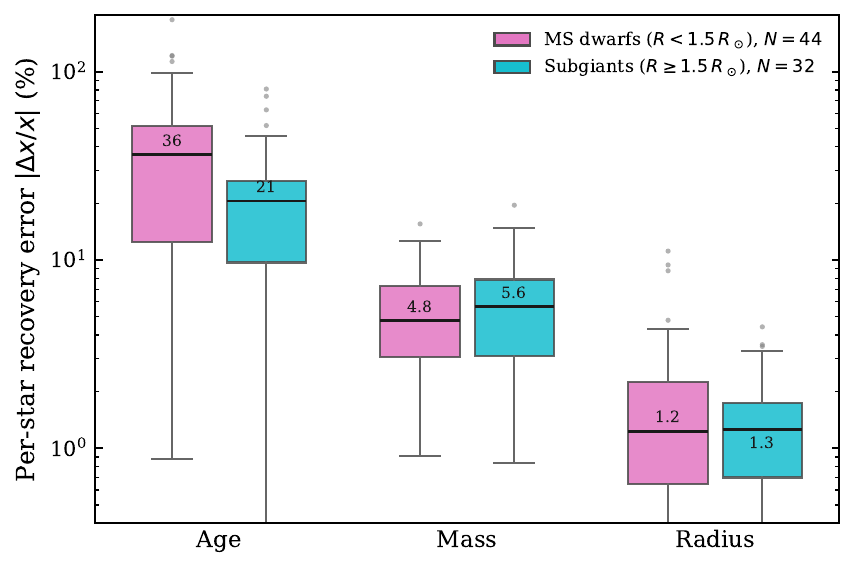}
    \caption{Boxplot with the fractional recovery error for the Combined-prior asteroseismic sample, split at $R=1.5R_\odot$ into main sequence (pink) and subgiants (cyan). The age error roughly halves past the turnoff while mass and radius are largely unchanged.}
    \label{fig:regime}
\end{figure}

The age recovery strongly depends on where a star sits in its evolution. Splitting the asteroseismic sample at $R=1.5R_\odot$ into main sequence dwarfs ($N=44$) and subgiants ($N=32$), the median absolute age error is $37\%$ on the main sequence but only $19\%$ on the subgiant branch (Figure~\ref{fig:regime}). This is expected: on the main sequence the isochrones are nearly age-degenerate, a dwarf's position in the $T_{\rm eff}-$luminosity plane barely shifts as it ages, so the photometry pins age weakly; past the turnoff the isochrones spread and age becomes recoverable. The bias is comparable in both regimes ($\sim5\%$ on the main sequence and $\sim8\%$ on the subgiant branch), so the $\sim22\%$ ensemble scatter is an average over a strong regime dependence, and a \textsc{lachesis} age should be trusted far more for an evolved star than for a dwarf. Mass and radius are insensitive to the split (radius $1.2\%$ vs $1.4\%$, mass $4.7\%$ vs $5.5\%$ median error), as expected for quantities fixed by luminosity and effective temperature rather than the interior clock. The evolutionary state itself is recovered for $74/76$ stars ($97\%$); the two errors are main sequence stars within $0.05$ dex of the $\log~g=4.0$ boundary.

\subsection{Photometric uncertainties and error calibration}
\label{sec:errors}

The fitted photometric excess-noise term (Section~\ref{sec:methods}) doubles as a measurement of how far catalogue broad-band uncertainties can be trusted. Its median across the photometric fits is $0.028$ mag against a typical catalogue per-band error of $0.020$ mag, an effective uncertainty $\approx1.7\times$ the quoted value, with $45\%$ of stars requiring more than $0.03$ mag and a tail to $\sim0.4$ mag where the photometry is discrepant. The effect on calibration is direct. Fitting without jitter results in a radius pull distribution width of $1.77$, badly over-confident, the radii missing the asteroseismic truth by nearly twice their stated errors. In contrast, fitting with jitter brings the radius pull distribution width to $1.04$, mass to $1.21$, and age to $1.01$. The radius and age pulls sit at essentially unit width, while the mass pull is slightly broad at $1.21$ with a median of $-0.7$, so the mass uncertainties remain mildly underestimated and retain a residual sub-$\sigma$ offset that the jitter term does not fully absorb.

We test the calibration of the reported posteriors with the PIT and the pull. Treating the asteroseismic values as exact, the internal $68\%$ credible intervals under-cover for mass and radius (empirical coverage $0.51$ PIT non uniform at $p<10^{-3}$), meaning the intervals are formally overconfident. This test is however unfair to the code because it ignores the asteroseismic error budget. Once the reference $1\sigma$ is folded in, the pull widths fall to near unity for all three parameters (Table~\ref{tab:calib}), consistent with a complete joint error budget. The residual signal is a location offset rather than a width miscalibration: the Combined-prior mass posterior carries a $-3.4\%$ ($\sim0.7\sigma$) systematic bias whose bootstrap confidence interval excludes zero. Age and radius biases are consistent with zero within their bootstrap intervals.

\subsection{Averaging versus model selection}
\label{sec:averaging}
The value of the model average is not a smaller point error, but a posterior that carries the systematic uncertainty of the grid choice. We make this explicit by comparing, on the 76 asteroseismic benchmarks, three ways of combining the grids: the evidence-weighted average used throughout (BMA), selection of the single highest-evidence grid per star, and each of the five grids used alone. Table~\ref{tab:averaging} reports the bias, RMSE, posterior widths, and empirical coverage of each.

\begin{table*}
\centering
\caption{Combining the grids on the $76$ asteroseismic benchmarks.}
\label{tab:averaging}
\begin{tabular}{lcccc}
\hline\hline
Scheme & Bias (Gyr) & RMSE (Gyr) & $W_{68}$ (Gyr) & $C_{68}$ / $C_{95}$ \\
\hline
\multicolumn{5}{l}{Combined-prior, age}\\
BMA (evidence)      & $+0.22$ & $2.48$ & $2.34$ & $0.64 / 0.89$ \\
Best-grid selection & $+0.07$ & $2.76$ & $2.16$ & $0.51 / 0.80$ \\
MIST only           & $+0.03$ & $2.69$ & $2.32$ & $0.61 / 0.88$ \\
PARSEC only         & $-0.15$ & $2.43$ & $2.08$ & $0.59 / 0.81$ \\
Dartmouth only      & $+0.44$ & $2.26$ & $2.35$ & $0.59 / 0.79$ \\
BaSTI only          & $+0.43$ & $2.28$ & $2.17$ & $0.59 / 0.83$ \\
YaPSI only          & $+0.25$ & $2.40$ & $2.18$ & $0.54 / 0.83$ \\
\hline
\multicolumn{5}{l}{No-prior, age}\\
BMA (evidence)      & $-0.78$ & $2.67$ & $3.47$ & $0.76 / 0.97$ \\
Best-grid selection & $-0.92$ & $2.95$ & $2.80$ & $0.59 / 0.84$ \\
MIST only           & $-1.20$ & $3.24$ & $3.83$ & $0.66 / 0.95$ \\
PARSEC only         & $-1.32$ & $2.94$ & $2.67$ & $0.61 / 0.80$ \\
Dartmouth only      & $+0.03$ & $2.45$ & $3.34$ & $0.68 / 0.89$ \\
BaSTI only          & $+0.23$ & $2.68$ & $3.13$ & $0.64 / 0.84$ \\
YaPSI only          & $-0.53$ & $2.38$ & $2.91$ & $0.71 / 0.92$ \\
\hline\hline
\end{tabular}
\tablefoot{
Columns give the signed bias and RMSE (physical units), the median $68\%$ credible-interval width $W_{68}$, and the empirical $68\%/95\%$ coverage, for age (Combined-prior above, No-prior below). Model averaging beats best-grid selection on accuracy and calibration, and the averaged posterior is the best-calibrated of the schemes tested; selection and the single grids reach a narrower $W_{68}$ but under-cover, so their sharper posteriors are overconfident.
}
\end{table*}

Two results follow. First, averaging is more accurate and better calibrated than selecting the single best-evidence grid, which is the worst or near-worst on RMSE and the most strongly under-covering (age coverage $0.51$ Combined-prior, $0.59$ No-prior). The single most accurate grid is moreover not knowable in advance and changes with the parameter and prior set, from Dartmouth for Combined-prior age to YaPSI for No-prior age to BaSTI for Combined-prior mass, so no fixed choice avoids the worst case that averaging bounds. Second, averaging significantly improves the age coverage relative to selecting the single highest-evidence grid. The averaged $68\%$ coverage is $0.64$ against $0.51$ for selection in the Combined-prior case, a paired-bootstrap difference of $+0.13$ ($95\%$ CI $[0.04, 0.22]$), and $0.76$ against $0.59$ in the No-prior case ($+0.17$, $[0.08, 0.26]$). Against any individual grid the averaged coverage is equal or higher ($0.54$--$0.61$ Combined-prior), but at $N=76$ these per-grid differences are not individually significant, so averaging gives the best calibration available without knowing in advance which grid to trust. Dropping any one grid changes the averaged coverage by at most $0.07$ (leave-one-grid-out), so the result is not driven by a single grid.

The width column shows where the calibration comes from. Selecting one grid, or trusting a single grid, gives a narrower posterior than the average (No-prior $W_{68}=2.80$ versus $3.47$~Gyr), but that sharpness is overconfidence, since those schemes under-cover. The average is wider by the grid-to-grid systematic it restores, and that is what makes its intervals contain the truth. Broadening every posterior sample by the asteroseismic reference error preserves this ordering, so the calibration advantage is not an artefact of neglecting the reference uncertainty. This is the sense in which model averaging improves the inference by making the reported uncertainty honest.

We verified that the inferred parameters and the model weights are insensitive to the choice of IMF, photometric-jitter, and extinction priors. Reweighting the Combined-prior posteriors of the 76 asteroseismic stars from the Chabrier IMF to the Salpeter or Kroupa forms, and to a tighter jitter ($\sigma_j\leq0.1$~mag) or extinction ($A_V\leq0.3$) prior, shifts the recovered masses and ages by less than $0.1\%$ and the mean model weights by less than $0.01$. The insensitivity to the IMF follows because the benchmark spans $0.8-1.6\,M_\odot$, above the sub-solar regime where the three IMFs diverge, and the jitter and extinction posteriors already lie within the tightened bounds.

\subsection{Injection-recovery and out-of-ensemble robustness}
\label{sec:injection}

To test directly whether the averaged posterior covers the truth, and in particular whether it remains honest when the generating physics is not one of our five grids, we forward-generate synthetic stars and fit them with the full pipeline. Each star is drawn as an FGK dwarf or subgiant ($5000<T_{\rm eff}<6800$~K, $3.4<\log~g<4.5$, $0.8<M<1.6M_\odot$), placed at 50~pc with zero extinction, and its broad-band magnitudes are computed through the same bolometric-correction table used in the fit and perturbed by 0.02~mag of Gaussian noise. We generate from MIST, which is inside the ensemble, as a calibration check, and from the non-rotating Geneva grid of \citet{mowlavi12}, which is a real independent evolution code not among our five, as an out-of-ensemble misspecification test. Each synthetic star is fit with the five-grid model average under a Gaussian [Fe/H] prior defined by the true value. Table~\ref{tab:injection} reports the recovery and the empirical coverage of the injected mass, radius, and age over 40 stars per source grid.

\begin{table}
\centering
\caption{Injection-recovery over $40$ synthetic FGK stars per source grid.}
\label{tab:injection}
\begin{tabular}{llcccc}
\hline\hline
Source & Param. & Bias (\%) & RMSE (\%) & $C_{68}$ & $C_{95}$ \\
\hline
MIST  & Mass   & $+3.3$  & $5.5$  & $0.75$ & $0.98$ \\
                     & Radius & $-1.1$  & $1.8$  & $0.82$ & $1.00$ \\
                     & Age    & $-15.7$ & $30.3$ & $0.75$ & $0.98$ \\
\hline
Geneva & Mass   & $+1.2$  & $5.0$  & $0.82$ & $1.00$ \\
                     & Radius & $-1.2$  & $2.0$  & $0.75$ & $0.98$ \\
                     & Age    & $-22.0$ & $33.3$ & $0.75$ & $0.98$ \\
\hline\hline
\end{tabular}
\tablefoot{
Bias is the median fractional residual, RMSE the full fractional scatter, and $C_{68}$/$C_{95}$ the empirical coverage of the $68\%$ and $95\%$ credible intervals of the five-grid average. Coverage is consistent with nominal within the $N=40$ binomial uncertainty for both the in-ensemble (MIST) and out-of-ensemble (Geneva) generating models.
}
\end{table}

The averaged posterior covers the injected truth consistently with the nominal rate in this experiment for both generating models: the $68\%$ credible intervals contain the truth for $75$ to $82\%$ of stars and the $95\%$ intervals for $98$ to $100\%$, with no degradation from the in-ensemble to the out-of-ensemble case. The point estimates carry the expected systematics, a few per cent in mass and radius and a young age bias of $16$ to $22\%$ that is the main-sequence age degeneracy rather than a property of the averaging (a single grid-fit shows the same young bias). The bias appears even when the generating grid is MIST, which is inside the ensemble, and a MIST-only fit of the same stars returns it unchanged, confirming it is the photometric age degeneracy of unevolved stars. The out-of-ensemble age bias exceeds the in-ensemble value by the additional model mismatch, yet the coverage is unchanged, so the width of the averaged posterior absorbs the misspecification. This is the sense in which the model average is robust: the reported uncertainty covers the truth even for a generating model outside the ensemble. We note that Geneva is a standard non-rotating code, so this test bounds robustness to a grid outside the ensemble and not to arbitrary unmodelled physics.

\section{Discussion}
\label{sec:disc}

\textsc{lachesis} recovers stellar mass and radius comparably to the literature grid-modelling floors ($4.3\%$ and $1.6\%$ NMAD), automatically from little more than a Gaia source identifier, though the user can supply their own photometry. The radius is essentially unbiased, while the mass carries a small but real $-3.4\%$ systematic underestimate whose confidence interval excludes zero and deepens to $\sim5\%$ for the most massive, evolved stars.

Mass and radius rest on firmer ground than age, though not because we have an independent calibration of them. The dynamical masses of the eclipsing binaries are model-independent, but their blended photometry forces a spectroscopy-only fit that constrains mass only weakly, so the mass scale, like the ages, is ultimately tested against the model-based asteroseismic reference. What sets mass and radius apart is that they are fixed almost directly by the luminosity and effective temperature, whereas age is the interior clock that broad-band photometry barely reads. The age recovery is reliable only for evolved stars and floor-limited for main-sequence dwarfs (Section~\ref{sec:regime}).

The inter-grid scatter of Section~\ref{sec:floor} is a model-dependent term that a single-grid posterior hides, subdominant to the data scatter for mass and radius and a $\sim9\%$ floor, about a third of the total width, for age. Carrying all five grids and weighting them by evidence exposes it and folds it into the reported uncertainty. The evidence weights should not be read as a ranking of physical accuracy. BaSTI returns the most accurate masses and radii of the five yet is only modestly weighted, because the evidence measures the fit to the photometric surface on a common scale rather than the accuracy of the derived parameters, and the near-zero age bias of MIST is coincidental, inverting to $-31\%$ once the metallicity prior is removed. On this benchmark the weights are statistically indistinguishable from flat, which we read as confirmation that no grid is pathologically favoured rather than as a shortcoming. The value of the average is therefore a better-calibrated posterior rather than a sharper one. It is more accurate and better calibrated than selecting the single highest-evidence grid, and the injection test confirms that its intervals cover the truth near the nominal rate even when the generating model lies outside the ensemble (Section~\ref{sec:injection}).

Several caveats bear on using \textsc{lachesis} in practice. Catalogue broad-band uncertainties are optimistic; the fitted jitter sits at a median $0.028$ mag against typical quoted errors of $0.020$ mag, and marginalising over it is what returns the parameter pulls to near-unit width (Section~\ref{sec:errors}), so a fit that takes the catalogue errors at face value reports uncertainties that are too small. A spectroscopic metallicity prior buys precision, not accuracy, tightening the posteriors by tens of per cent while leaving the median recovery statistically unchanged (Section~\ref{sec:recovery}). The photometric radius assumes a single star; an unresolved companion inflates it by up to $\sim30\%$ in proportion to its light, so an astrometric binarity screen such as a Gaia \textsc{ruwe} cut should precede any single-star fit. The inferred parameters and weights are, moreover, insensitive to the IMF, jitter, and extinction priors over the mass range of the benchmark (Section~\ref{sec:averaging}).

Two limitations of the grid set remain. BaSTI is interpolated on a mass-sorted index that clips the intermediate-mass convective-core hook, adding $\sim1\%$ in temperature for the $\sim1.5-2.5M_\odot$ turnoff stars in the sample (Appendix~\ref{sec:interp}), and because each grid adopts its own solar reference and [Fe/H]$\rightarrow Z$ mapping, a common numerical metallicity spans $\sim0.06$ dex in physical $Z$. Both enter the inter-grid systematic that the model-averaged uncertainty reflects.

A natural extension to this framework and means to probe the interior clock is through incorporating asteroseismic constraints. At the simplest level, global seismic parameters such as $\Delta\nu$ and $\nu_{\max}$ could be added by using grids in which these quantities have already been tabulated \citep[e.g., BaSTI,][]{hidalgo_updated_2018} or by augmenting existing isochrone grids with seismic predictions. Bayesian grid-based modelling of asteroseismic targets has already demonstrated the value of this approach, particularly when simple scaling-relation estimates of global seismic parameters are replaced by model-based values computed with oscillation codes \citep[e.g.,][]{rodrigues2017,basta}. The motivation for doing this in a BMA framework is not only to increase precision but also to place the systematic uncertainty budget on the same footing as the statistical one. This is particularly important for high-priority Habitable Worlds Observatory target stars, for which precise masses and ages are needed to inform target prioritisation and mission trade studies \citep[e.g.,][]{tuchow2022,harada2024,tuchow2025,ware2025}, interpret planetary system architectures, assess the dynamical viability of habitable zones \citep{kane2024}, and place potentially habitable planets in an evolutionary context \citep{bixel2020,unterborn2022}. Even for these comparatively well-studied stars, the systematic uncertainties reported for asteroseismic ages are comparable to the formal statistical uncertainties ($\sim 10-20\%$), despite asteroseismology currently being the most precise age-dating method available for many nearby field stars \citep{hages}. Previous work has shown that asteroseismic inferences can be limited by assumptions about stellar input physics and seismic modelling choices, including the treatment of $\Delta\nu$ corrections, initial helium abundance, mixing length, diffusion, overshooting, and evolutionary state \citep[e.g.,][]{valle2015,tayar2022,li2024,tayar2025}. Within a BMA framework, the inference would therefore not be tied to a single set of stellar models or seismic corrections.

A more ambitious extension would be to include individual oscillation frequencies, frequency ratios, or separation ratios in the likelihood. This would require each stellar model grid to be supplemented with adiabatic oscillation calculations, together with a consistent treatment of surface effects, mode selection, and interpolation across evolutionary tracks. Although this is computationally more demanding than fitting global seismic parameters, it is feasible and is already the basis of detailed asteroseismic pipelines such as AIMS \citep{aims} and BASTA \citep{basta}. The benefit would be greatest for main sequence and subgiant stars, for which individual frequencies can provide substantially tighter age constraints than classical isochrone fitting alone \citep{silvaaguirre17}. However, the same precision that makes individual-frequency modelling powerful also makes it more sensitive to model deficiencies, surface corrections, and differences in the adopted stellar physics. Therefore, a BMA treatment would be especially valuable in this regime.

\section{Conclusions}
\label{sec:conc}

We have presented \textsc{lachesis}, a \textsc{python} package that infers stellar masses, radii, and ages by fitting an ensemble of isochrone grids to archival photometry and spectroscopy and combining the per-grid results through Bayesian model averaging. Each grid is interpolated in [Fe/H], log-age, and equivalent evolutionary phase, sampled with nested sampling, and weighted by its Bayesian evidence, so that the systematic uncertainty arising from the choice of stellar evolution model is propagated into a single posterior on the stellar parameters rather than carried as a separate error term.

We validated the code against a benchmark of Kepler asteroseismic dwarfs and subgiants and a sample of detached eclipsing binaries with model-independent dynamical masses and radii. Our main results are as follows.

\begin{itemize}
    \item \textsc{lachesis} recovers stellar masses to a median accuracy of $\sim4\%$ and radii to $\sim2\%$, comparable to the literature model floors; the radius is unbiased and the mass carries a small $\sim3\%$ systematic underestimate.
    \item The age uncertainty is dominated by the weak photometric leverage on the interior clock; grid choice adds a real but subdominant floor of $\sim9\%$ (about $11\%$ of the total age variance). The robust scatter is $\sim30\%$ and depends strongly on the evolutionary state; ages are reliable beyond the turnoff, but degenerate on the main sequence.
    \item The five grids agree on age at the $\sim9\%$ level at fixed input, which we interpret as a lower bound on the model-dependent age floor reported in the literature \citep{tayar2022}. A variance decomposition shows this inter-grid term is $\sim11\%$ of the total age variance, so it is a real but subdominant contribution that single-grid fits omit.
    \item The averaged posterior is the best-calibrated of the combination schemes: it is better calibrated than any single grid and than selecting the highest-evidence grid. Its real-benchmark credible intervals under-cover when the reference is treated as exact ($68\%$ coverage $0.51$--$0.64$) and approach consistency once the reference uncertainty is included; in the injection test they cover the truth near the nominal rate even when the generating model lies outside the ensemble. Folding the grid-choice systematic into an honest, calibrated uncertainty is the central advantage of the method over using or selecting a single grid.
\end{itemize}

\noindent The evidence weights reduce to near-flat weights on this benchmark, confirming that the grids are comparably supported; the value of the method lies in the calibration and robustness of the averaged posterior rather than in a sharper point estimate.

Several extensions follow naturally. The framework is agnostic to the number and identity of the grids, so the ensemble can be expanded with additional model grids that cover a wider range of input physics, for example, those that include rotation \citep[e.g.,][]{syclist,starevol,parsecv2} or incorporate more realistic prescriptions for convective overshoot \citep{young05}. Models that introduce extra structural dimensions, such as the radius-inflating effect of star spots \citep{somers20}, can enter as additional nuisance axes, letting the evidence quantify star by star when that physics is warranted. Finally, because \textsc{lachesis} consumes the same posterior products as the SED-fitting code \textsc{ariadne} \citep{vines_span_2022}, a joint treatment of the photometric, spectroscopic, and isochrone constraints is the natural next step towards homogeneous stellar parameters for the large samples of exoplanet hosts that forthcoming surveys will deliver.

\begin{acknowledgements}
We thank the anonymous reviewer for their insightful comments that improved the content of this manuscript. José Vines acknowledges financial support from ANID / Fondo ALMA 2024 / No. 31240039 “On the origin of Warm-Giant Planets.” Austin Ware acknowledges support from NASA Grant No. 80NSSC23K1356 (PI: Steve Desch) issued through the Science Mission Directorate Planetary Science Division.
\end{acknowledgements}

  \bibliographystyle{aa}
  \bibliography{references}

\begin{appendix}
\nolinenumbers
\section{Validation of the interpolation coordinate}
\label{sec:interp}

Because BaSTI is interpolated on its native mass-sorted index rather than on physics-based EEPs, we test directly whether that index is an adequate interpolation coordinate rather than relying on its sampling density. We remove one age node, and separately one metallicity node, and reconstruct it by linear interpolation from its two neighbours at fixed index, the operation the interpolation performs when a star falls between grid nodes. Across the age axis the BaSTI index is excellent: the held-out fractional error at the main-sequence turnoff is $0.01\%$ in effective temperature and $0.15\%$ in luminosity, even though a fixed index maps to a very different stellar mass at different ages (from $0.5$ to $1.4M_\odot$ for a single interior index). Across metallicity, the index is weakest but still modest, rising to $0.5\%$ in temperature and $2.4\%$ in luminosity at the turnoff. For the field dwarfs, subgiants and old turnoff stars that dominate our sample, the mass index, the reconstructed EEP, and the phase-based EEP are interchangeable at the few-per cent level in the median; the 95th-percentile luminosity error reaches $\sim18\%$ on the fastest-evolving nodes (Table~\ref{tab:interp}), so the interpolation is typically accurate to a few per cent but not uniformly.

\begin{table}[!ht]
\centering
\caption{Held-out interpolation error of the grid coordinate at the main-sequence turnoff.}
\label{tab:interp}
\begin{tabular}{llcc}
\hline\hline
Grid (coordinate) & Axis & $T_{\rm eff}$ (\%) & $L$ (\%) \\
\hline
BaSTI (mass index)     & age    & $0.01 / 0.63$ & $0.15 / 7.7$ \\
MIST (phase EEP)       & age    & $0.06 / 0.37$ & $0.80 / 2.4$ \\
Dartmouth (phase EEP)  & age    & $0.04 / 1.14$ & $0.26 / 13.4$ \\
YaPSI (reconstr.\ EEP) & age    & $0.10 / 0.94$ & $1.29 / 17.7$ \\
\hline
BaSTI (mass index)     & [Fe/H] & $0.51 / 1.99$ & $2.42 / 11.8$ \\
MIST (phase EEP)       & [Fe/H] & $0.32 / 1.90$ & $0.64 / 5.0$ \\
Dartmouth (phase EEP)  & [Fe/H] & $0.30 / 1.39$ & $1.59 / 10.2$ \\
YaPSI (reconstr.\ EEP) & [Fe/H] & $0.05 / 0.29$ & $0.39 / 2.9$ \\
\hline\hline
\end{tabular}
\tablefoot{
Values are the fractional error (median / 95th percentile, per cent) in recovering a dropped node from its two neighbours by interpolation across the age and metallicity axes at fixed index. The BaSTI mass index is the most accurate across age and the weakest across metallicity. The medians agree at the few-per-cent level; the 95th-percentile luminosity errors (up to $\sim18\%$) show the agreement is not uniform.
}
\end{table}

The regime in which the mass index fails is the intermediate-mass convective-core hook ($\sim1.5$ to $2.5M_\odot$, ages $\sim0.5$ to $2$~Gyr), where the coordinate compresses the entire rise to the turnoff into a mass interval of $\sim10^{-3}M_\odot$ and linear interpolation clips the sharp extremum. There, the interpolation error reaches $\sim1.1\%$ in effective temperature and $\sim2.7\%$ in luminosity even at BaSTI's native resolution, and it does not improve with denser sampling because it is set by the curvature of the corner rather than the node count. The phase-based MIST EEP renders the same feature as a smooth dome and is roughly 25 times more accurate there ($0.04\%$ and $0.17\%$); the central-hydrogen-anchored YaPSI EEP is intermediate. 13 of our 76 asteroseismic benchmark stars (six above $1.5M_\odot$, up to $1.6M_\odot$) fall in this regime, so we treat the mass-index interpolation as an additional model uncertainty of order $1\%$ in temperature and a few per cent in luminosity for intermediate-mass turnoff stars. Its effect on the inferred parameters is smaller than the point-wise clip, because the likelihood integrates over the evolutionary coordinate and BaSTI is one of five averaged grids; on the affected stars, BaSTI's age recovery departs from the ensemble by only $\sim3\%$.

\section{Photometric quality control}
\label{sec:qc}

\begin{figure}[!ht]
    \centering
    \begin{subfigure}{0.95\linewidth}
        \centering
        \includegraphics[width=\linewidth]{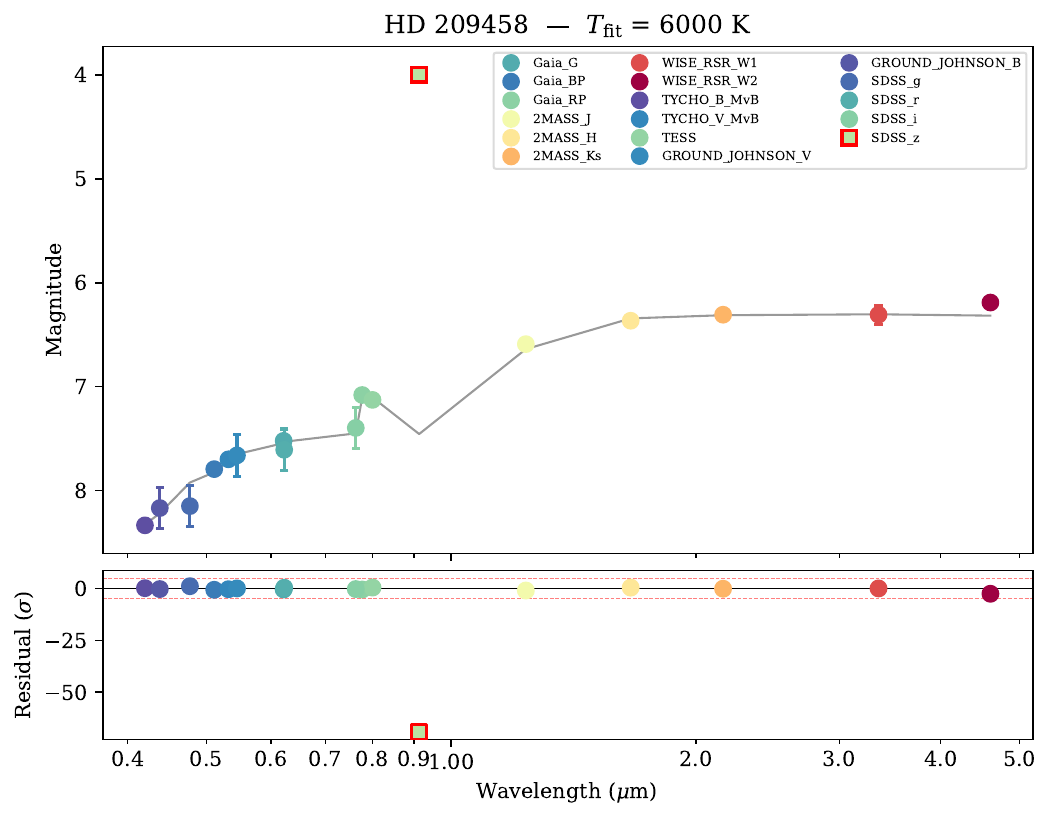}
        \caption{}
        \label{fig:bbody}
    \end{subfigure}

    \begin{subfigure}{0.95\linewidth}
        \centering
        \includegraphics[width=\linewidth]{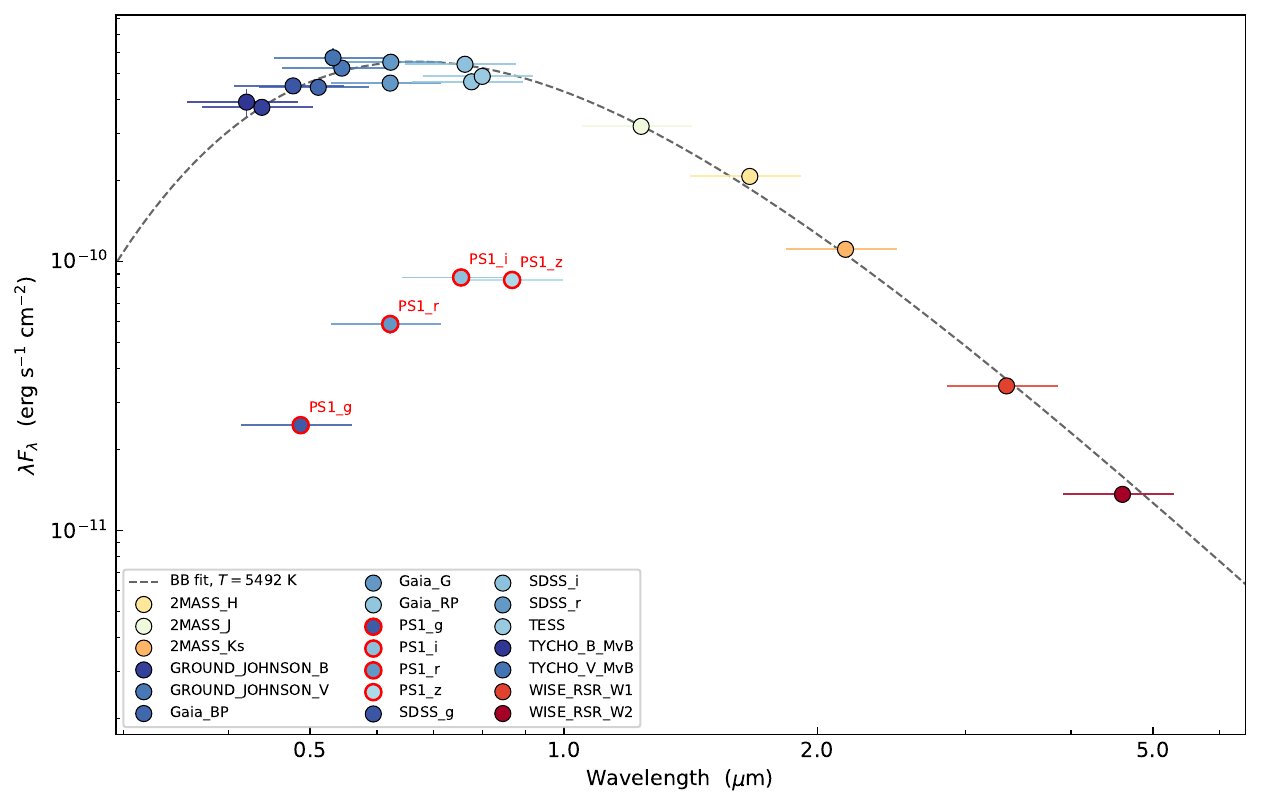}
        \caption{}
        \label{fig:bbody2}
    \end{subfigure}
    \caption{Photometric quality control. (a) HD~209458, with an artificially injected outlier in the SDSS $z$ band (red square, $m_z = 4.0$) to illustrate the procedure. Top: Observed magnitudes (circles with error bars) compared to the best-fit synthetic model at $T_{\rm fit} = 6000$~K (grey line), computed from the MIST bolometric correction tables. The model is anchored on Gaia photometry and iteratively extended to include all consistent bands. Bottom: Normalised residuals in units of the photometric uncertainty. Dashed lines mark the $\pm 5\sigma$ rejection threshold. Only the injected measurement is flagged ($69\sigma$). (b) TYC~6296-96-1, illustrating the blackbody-fit outlier detection offered as an SED plot for visual inspection.}
    \label{fig:qc}
\end{figure}

Figure~\ref{fig:qc} illustrates the photometric quality control of Section~\ref{sec:phot}.

\end{appendix}

\end{document}